\pdfoutput=1
\documentclass[11pt]{article}
\usepackage[]{acl}

\usepackage[T1]{fontenc}
\usepackage{times}
\usepackage{latexsym}
\usepackage{enumitem}
\usepackage{color, colortbl}
\usepackage{threeparttable,booktabs}
\usepackage{amssymb}
\usepackage{graphicx,subcaption}
\usepackage{amssymb}
\usepackage{xurl}
\usepackage[hang,flushmargin]{footmisc}

\usepackage{microtype}
\usepackage{booktabs}
\usepackage{multirow}
\usepackage{adjustbox}
\usepackage{arydshln}
\usepackage{footnote}
\usepackage{tabularx}
\usepackage{array}
\usepackage{graphicx,subcaption}
\usepackage{enumitem}
\usepackage{amsmath}
\usepackage{amsfonts}
\usepackage{dsfont}
\usepackage{pifont}
\usepackage{microtype}

\definecolor{bluencs}{rgb}{0.0, 0.53, 0.74}
\newcommand{\cmark}{\ding{51}}%
\newcommand{\xmark}{\ding{55}}%

\definecolor{Gray}{gray}{0.9}

\newcommand{\ourmodel}{\textsc{Dragon}}

\title{How to Train Your \ourmodel{}:\\Diverse Augmentation Towards Generalizable Dense Retrieval
}

\author{Sheng-Chieh Lin$^1$\thanks{\ \ This work is done during Sheng-Chieh's internship at Meta.}\ , 
Akari Asai$^2$,
Minghan Li$^1$,
Barlas Oguz$^3$,\\
{\bf Jimmy Lin$^1$},
{\bf Yashar Mehdad$^3$},
{\bf Wen-tau Yih$^3$}, \and
{\bf Xilun Chen$^3$\thanks{\ \ Xilun and Sheng-Chieh contributed equally to this work.}}\\[1ex]
        University of Waterloo$^1$, University of Washington$^2$, Meta AI$^3$\\[1ex]
        \texttt{\{s269lin,m692li,jimmylin\}@uwaterloo.ca}, 
        \texttt{akari@cs.washington.edu}\\
        \texttt{\{barlaso,mehdad,scottyih,xilun\}@meta.com}\\
}

\begin{document}
\maketitle
\begin{abstract}
Various techniques have been developed in recent years to improve dense retrieval (DR), such as unsupervised contrastive learning and pseudo-query generation. 
Existing DRs, however, often suffer from effectiveness tradeoffs between supervised and zero-shot retrieval, which some argue was due to the limited model capacity.
We contradict this hypothesis and show that a generalizable DR can be trained to achieve high accuracy in both supervised and zero-shot retrieval without increasing model size.
In particular, we systematically examine the contrastive learning of DRs, under the framework of Data Augmentation (DA).
Our study shows that common DA practices such as query augmentation with generative models and pseudo-relevance label creation using a cross-encoder, are often inefficient and sub-optimal.
We hence propose a new DA approach with diverse queries and sources of supervision to progressively train a generalizable DR.
As a result, \ourmodel,\footnote{The code and model checkpoints are available at: \url{https://github.com/facebookresearch/dpr-scale}} our \textbf{D}ense \textbf{R}etriever trained with diverse  \textbf{A}u\textbf{G}mentati\textbf{ON}, is the first BERT-base-sized DR to achieve state-of-the-art effectiveness in both supervised and zero-shot evaluations and even competes with models using more complex late interaction (ColBERTv2 and SPLADE++).

\end{abstract}

\maketitle

\section{Introduction}
\label{sec:intro}
\begin{figure}[t]
    \centering
    \resizebox{1\columnwidth}{!}{
        \includegraphics{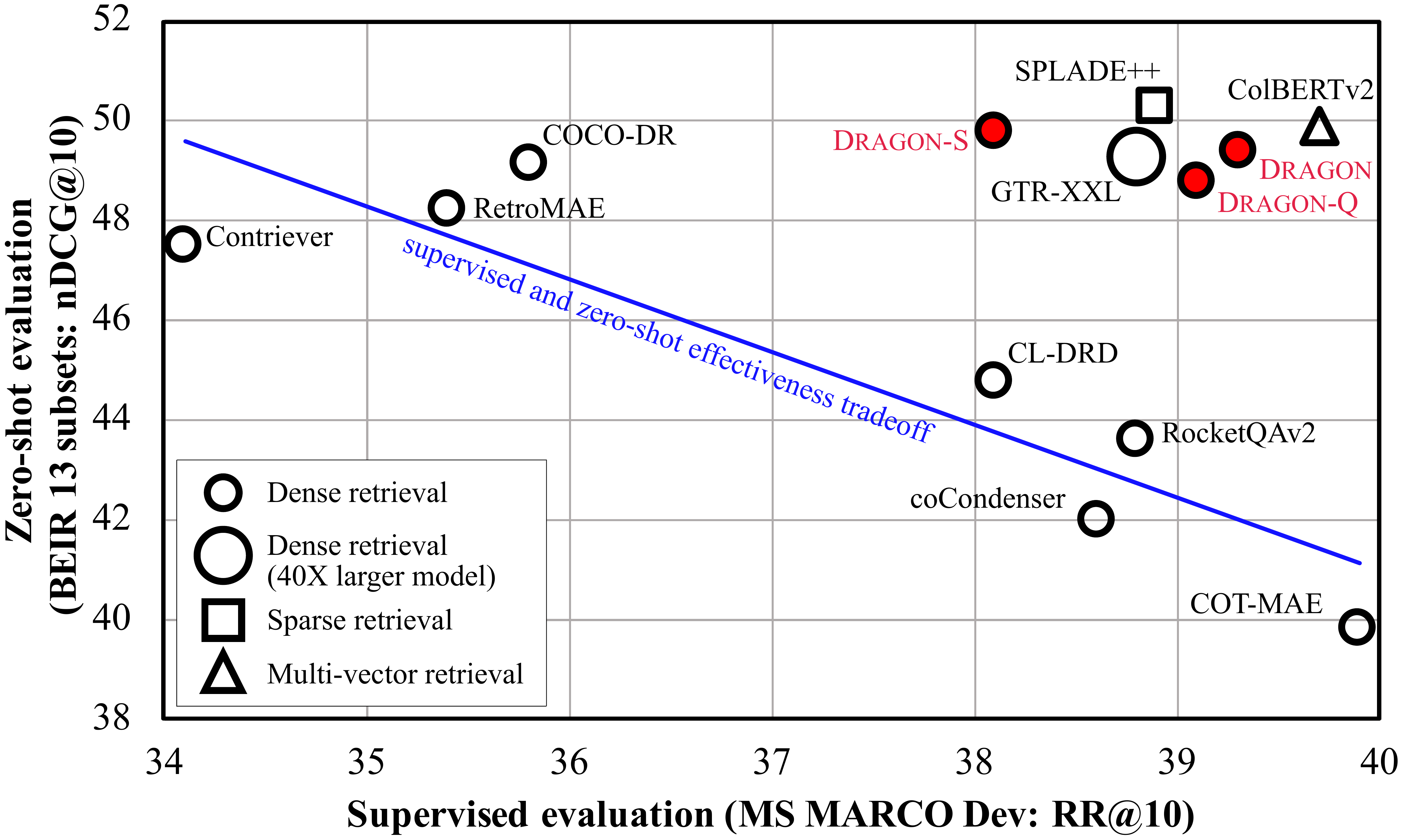}
    }
    \caption{Supervised versus zero-shot effectiveness comparison among existing state-of-the-art retrievers. 
    % All models use a BERT-base-sized (110M parameters) backbone except for GTR-XXL (4.8B parameters).
    }  
    \label{fig:teaser}
\end{figure}

Bi-encoder based neural retrievers allow documents to be pre-computed independently of
queries and stored, enabling end-to-end retrieval among huge corpus for downstream knowledge-intensive tasks~\citep{dpr, sentence-bert}. 
Recently, \citet{beir} show that it is challenging to deploy such bi-encoder retrievers in real-world scenarios, where training data is scarce. 
One potential solution is to design more expressive representations to capture more fine-grained token-level information; e.g., SPLADE++~\cite{distil-splade} and ColBERTv2~\cite{colbert-v2} in Figure~\ref{fig:teaser}. 
However, these designs add complexity and latency to retrieval systems~\cite{Mackenzie_etal_arXiv2021}.

By contrast, dense retrieval (DR) is a simpler bi-encoder retrieval model, which maps queries and documents into low-dimensional vectors and computes text similarity through a simple dot product. 
Top-$k$ retrieval can be performed directly
using ANN search libraries~\cite{faiss, scann}. 
Recently, various methods have been proposed to improve DR effectiveness while keeping its simple architecture, such as pre-training~\cite{ict,Chang2020Pre-training}, query augmentation~\cite{domain-match-pretraining}, and distillation~\cite{rocketqav2, cldrd}. 
Some of these methods are very effective at improving accuracy in the fully supervised setting, while others can improve transfer in the zero-shot setting. 
In most cases, however, improving the former is only achieved at the expense of the latter.  
Figure~\ref{fig:teaser} plots existing state-of-the-art DR models with respect to their effectiveness on these two axes, which presents a clear tradeoff between supervised and zero-shot effectiveness (the blue line).  
The only exception, GTR-XXL~\cite{gtr}, breaks the effectiveness tradeoff at the expense of efficiency (i.e., query encoding), which leverages very large pre-trained models with 40 times more parameters. 
This effectiveness tradeoff prompts some to hypothesize that we have fully exploited the capacity of BERT-base-sized DR model~\cite{gtr} and explore how to cleverly increase model parameters without sacrificing retrieval efficiency.   
For example, the recent work~\cite{gpl, promptgator} proposes to train one expert dense retriever for each specific scenario, resulting in slow adaptation to real-world applications~\cite{tart}.

In this work, we contradict this hypothesis and show that a generalizable DR can indeed be trained to achieve state-of-the-art effectiveness in both supervised and zero-shot evaluations \textit{without} increasing model size. 
To this end, we first investigate the important factors contributing to the recent progress of DR. 
For example, DR seems to gain zero-shot transfer capability from pre-training on large-scale and diverse training queries~\cite{contriever, cocodr} while knowledge distillation can improve the supervision quality by automatically identifying relevant passages which are not labeled by humans~\cite{rocketqav2, cldrd}. 
To better understand these approaches, we devise a unified framework of data augmentation (DA) for contrastive learning. 
Under the framework, the previous work can be viewed as DA with different recipes of query augmentation and relevance label augmentation shown in Table~\ref{tb:model_comparison}.

Guided by a detailed empirical exploration along the space of our DA framework, we find the following:\ (1) 
for relevance label augmentation, we identify that the key to training a generalizable dense retriever is to create \emph{diverse} relevance labels for each query, for which we use multiple retrievers instead of a strong cross encoder; (2) with such diverse relevance labels, dense retrievers can be trained effectively using cheap and large-scale augmented queries (e.g., cropped sentences from a corpus) instead of the more expensive neural generative queries; see \ourmodel-S vs \ourmodel-Q in Figure~\ref{fig:teaser}. 
This finding opens the door to further building cheap but useful training data in scale for DR in the future. 
Finally, we find that it is suboptimal for a dense retriever to learn the diverse relevance labels from multiple retrievers directly. 
Thus, we propose a simple strategy to \emph{progressively} augment relevance labels which guides dense retrievers to learn diverse relevance signals more effectively. 

Our final model is trained on 28 million augmented queries consisting of two types (cropped sentences and synthetic queries), as well as progressive relevance label augmentation using diverse (sparse, dense, and multi-vector) retrievers.
As shown in Figure~\ref{fig:teaser}, \ourmodel, a \textbf{D}ense \textbf{R}etriever trained with diverse \textbf{A}u\textbf{G}mentati\textbf{ON}, is the \emph{first} dense retriever to break the supervised and zero-shot effectiveness tradeoff without increasing model size or retrieval complexity; e.g., GTR-XXL, SPLADE++ and ColBERTv2. 

We summarize our contributions as follows. 
\vspace{-0.2cm}
\begin{itemize}
\itemsep-0.4em 
\item We conduct a systematic study of DR training under the lens of data augmentation, which provides some surprising but key insights into training a generalizable dense retriever.   
\item We propose a progressive label augmentation strategy to guide a dense retriever to learn the diverse but complex relevance labels. 
\item \ourmodel, our single BERT-base-sized DR, reaches state-of-the-art retrieval effectiveness in both supervised (MS MARCO) and zero-shot (BEIR and LoTTE) evaluations. 
\end{itemize}

\begin{table}[t]
	\caption{Categorization of existing DR models by their approaches to data augmentation.}
	\label{tb:model_comparison}
	\centering
	 \resizebox{0.5\textwidth}{!}{  
    \begin{tabular}{ll!{\color{lightgray}\vrule}ccc}
	\hline \hline
 \multicolumn{1}{l}{Type}& \multicolumn{1}{l}{Model}& \multicolumn{1}{c}{Qry Aug}& \multicolumn{1}{c}{Label Aug}& \multicolumn{1}{c}{Corpus}\\ 
\hline
\arrayrulecolor{lightgray}
\multirow{2}{*}{1} & RocketQAv2& \multirow{2}{*}{\xmark}& \multirow{2}{*}{CE}& \multirow{2}{*}{MS MARCO}\\
 & CL-DRD& & & \\
 \hline
\multirow{3}{*}{2}& coCondenser& & & MS MARCO\\
& Contriever& cropping& \xmark& Wiki+CCnet\\
& COCO-DR& & & BEIR\\
\hline
\multirow{2}{*}{3} &GPL& \multirow{2}{*}{GenQ}& \multirow{2}{*}{\xmark}& \multirow{2}{*}{BEIR}\\
 &PTR& & & \\
 \arrayrulecolor{black}
\hline
& \ourmodel-S& cropping& & \\
& \ourmodel-Q& GenQ& retrievers& MS MARCO \\
& \ourmodel& cropping+GenQ& & \\
\arrayrulecolor{black}
	\hline \hline
	\end{tabular}
	}
\end{table} 

\section{Background}
In this section, we first introduce the retrieval task and contrastive learning approach for dense retrieval. 
We then provide a unified framework for understanding recent approaches to improve dense retrieval training as instances of Data Augmentation.

\subsection{Training Dense Retrieval Models}
Given a query $q$, our task is to retrieve a list of documents to maximize some ranking metrics such as nDCG or MRR.
Dense retrieval (DR) based on pre-trained transformers~\cite{devlin2018bert, raffel2019t5} encodes queries and documents as low dimensional vectors with a bi-encoder architecture and uses the dot product between the encoded vectors as the similarity score:
\begin{align}
\label{eq:cls_scoring_function}
    \textrm{s}(q, d) \triangleq  \mathbf{e}_{q_{\texttt{[CLS]}}} \cdot \mathbf{e}_{d_{\texttt{[CLS]}}},
\end{align}
\noindent where $\mathbf{e}_{q_{\texttt{[CLS]}}}$ and $\mathbf{e}_{d_{\texttt{[CLS]}}}$ are the $\texttt{[CLS]}$ vectors at the last layer of BERT~\citep{devlin2018bert}.

Contrastive Learning (CL) is a commonly used method for training DR models by contrasting positive pairs against negatives.
Specifically, given a query $q$ and its relevant document $d^+$, we minimize the InfoNCE loss:
\begin{align}
\label{eq:cl}
   -\log \frac{{\exp(\textrm{s}(q, d^+)})}{ \exp({\textrm{s}(q, d^+)}) + \overset{k}{\underset{j=1}{\sum}} \exp({\textrm{s}(q, d^-_j)})}. \end{align}
Eq.~(\ref{eq:cl}) is to increase the similarity score $\textrm{s}(q, d^+)$ and decrease the similarity scores between the query and the set of irrelevant documents $\{d_i^-\cdots d_k^-\}$.
The set of irrelevant documents, ideally, includes all the documents other than $d^+$ from the whole corpus, which, however, is not computationally tractable. 
Thus, an alternative is to mine hard negatives or cross-batch samples, which has been well studied by the previous work~\cite{xiong2020approximate, dpr, rocketqa}.

\subsection{A Unified Framework of Improved Dense Retrieval Training: Data Augmentation}
\label{subsec:augmentation_for_contrastive_learning}
Data augmentation (DA) for contrastive learning (CL) has been widely used in many machine learning tasks~\cite{simclr,augmented-sentencebert}. 
In fact, many recent approaches to train better DR, such as knowledge distillation, contrastive pre-training and pseudo query generation (GenQ), can be considered DA with different recipes respectively categorized as type 1, 2 and 3 in Table~\ref{tb:model_comparison}. 
We compare the different DA recipes from the perspectives of query and relevance label augmentation.

\smallskip \noindent \textbf{Query Augmentation.} 
There are two common automatic approaches to increase the size of training queries from a given corpus, sentence cropping and pseudo query generation. 
The former can easily scale up query size without any expensive computation, which are used by the type-2 models for contrastive pre-training~\cite{cocondenser, contriever, cotmae}.  
The latter generates quality but more expensive human-like queries using large language models for DR pre-training~\cite{domain-match-pretraining} or domain adaptation~\cite{gpl, promptgator}.

\smallskip \noindent \textbf{Supervised Label Augmentation.} 
The aforementioned approaches to query augmentation (i.e, types 2 and 3) often assume that the (or part of the) original document is relevant to the augmented queries, which may not be true and only provides a single view of relevance labeling. 
The recent work \citep[type 1;][]{rocketqav2, cldrd} improve DR training with the positive passages predicted by cross encoders. 
These approaches leverage knowledge distillation as an instance of label augmentation from human labels to improve data quality, inspiring us to further conduct supervised label augmentation on the augmented queries (i.e., cropped sentences and GenQ).

\subsection{Settings for Empirical Studies}
\label{subsec:datasets_and_evaluations}
We introduce some basic experimental settings to facilitate the presentation of our empirical studies on data augmentation in Section~\ref{subsec:training}. 
More detailed settings can be found in Section~\ref{sec:experiments}.
Following previous work~\cite{contriever, retromae, cocodr, distil-splade, colbert-v2}, we consider MS MARCO~\cite{marco} as supervised data and BEIR datasets for zero-shot evaluations. 
Thus, we use the 8.8 million MS MARCO passage corpus to conduct data augmentation and evaluate our trained models on MS MARCO Dev, consisting of 6980 queries from the development set with one relevant passage per query on average. 
We report MRR@10 (abbreviated as RR@10) and Recall@1000 (R@1K) as the evaluation metrics. 
For zero-shot evaluations, we use BEIR~\cite{beir}, consisting of 18 IR datasets spanning diverse domains and tasks including retrieval, question answering, fact checking, question paraphrasing, and citation prediction. 
We report the averaged nDCG@10 over 13 public BEIR datasets, named BEIR-13, making the numbers comparable to most existing approaches~\cite{splade-v2, colbert-v2}.\footnote{CQADupStack, Robust04, Signal-1M, TREC-NEWS, BioASQ are excluded}

\section{Pilot Studies on Data Augmentation}
\label{subsec:training}
In this section, we first discuss the exploration space of data augmentation (DA) based on the framework in Section~\ref{subsec:augmentation_for_contrastive_learning} and then conduct empirical studies on how to better train a dense retriever. 
Based on the empirical studies, we propose our DA recipe to train \ourmodel, a \textbf{D}ense \textbf{R}etriever with diverse \textbf{A}uGmentatiO\textbf{N}.

\subsection{An Exploration of Data Augmentation}
\smallskip \noindent \textbf{Query Augmentation.} 
% \subsubsection{Query Augmentation} 
Following the discussion in Section~\ref{subsec:augmentation_for_contrastive_learning}, we consider the two common approaches to automatic query augmentation.
Specifically, for sentence cropping, following \citet{spar}, we use the collection of 28 million sentences from the MS MARCO corpus consisting of 8.8 million passages. 
As for pseudo query generation, we use the 28 million synthetic queries sampled from the query pool generated by doct5query~\cite{doctttttquery}. 
In addition, we also consider augmenting the type of queries by mixing cropped sentences and synthetic queries. 

\smallskip \noindent \textbf{Label Augmentation with Diverse Supervisions.} 
Although cross encoder (CE) is known to create relevance labels with strong supervision, we hypothesize that CE still cannot capture diverse matching signals between text pairs. 
A query is often relevant to many documents from different perspectives (e.g., semantic or lexical matching), which cannot capture by a single labeling scheme (a strong model or even human).
In this work, we seek multiple sources of supervisions from existing sparse, dense and multi-vector retrievers, which are more efficient than CE and suitable for labeling a large number of queries (see discussion in Section~\ref{sec:discussion}).

\subsection{Training with Diverse Supervisions}
\label{subsec:supervsion_strategies}
We have introduced our searching space for query augmentation and label augmentation (with diverse supervisions); however, training a dense retriever on such augmented data is not trivial. 
First, how can we create supervised training data using a teacher from any augmented queries (i.e., cropped sentences or pseudo generative queries)? 
Second, with the supervised training data sets from multiple teachers, how can we train a dense retriever to digest the multiple supervisions?

Formally speaking, given $N$ teachers, for each augmented query $q$, we retrieve $N$ ranking lists (i.e., $\mathcal{P}_q^1, \mathcal{P}_q^2, \cdots, \mathcal{P}_q^N$ with each list has $K$ passages) from the corpus with the respective teachers. 
We consider the ranking list $\mathcal{P}_q^n$ from the $n$-th teacher a source of supervision since the top-$k$ and last-$k'$ passages in $\mathcal{P}_q^n$ contain the teacher's view on what is relevant and non-relevant for the given query. 
We then discuss possible strategies to train a dense retriever with diverse supervisions.

\begin{figure}[t]
    \centering
    \resizebox{0.98\columnwidth}{!}{
        \includegraphics{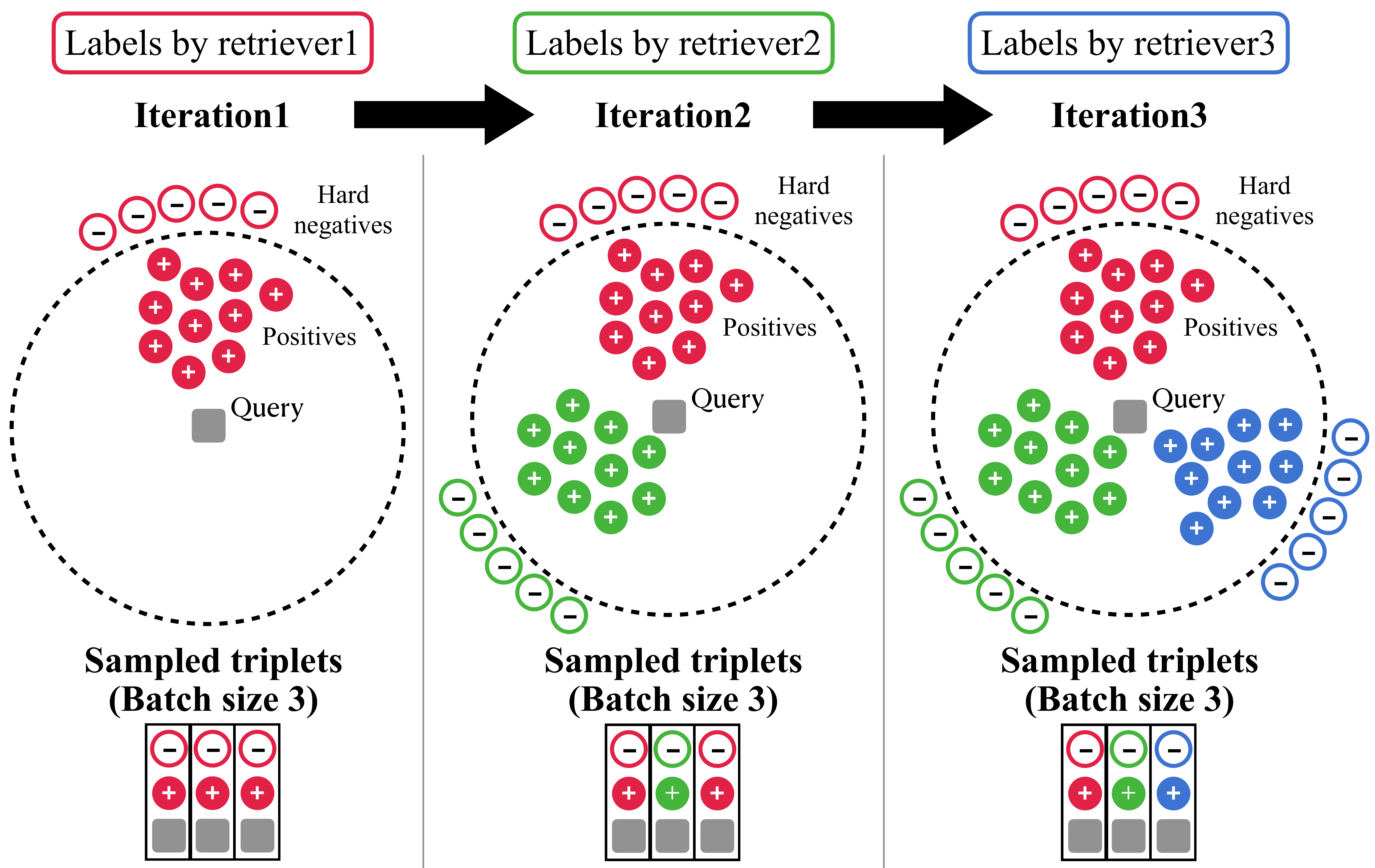}
    }
    \caption{Illustration of progressive label augmentation. For each iteration of training, additional relevance labels from a teacher are augmented in the training data. By contrast, uniform supervision directly exposes models to all the supervisions (as in iteration 3) in the beginning.} 
    \label{fig:approach}
\end{figure}

\smallskip \noindent \textbf{Fused Supervision.}
An intuitive strategy is to fuse the multiple sources into a single high-quality supervision, which dense retrievers can learn from. 
For the augmented query $q$, we conduct linear score fusion~\cite{dpr_repro} on the $N$ ranking lists to form a new ranking list $\mathcal{F}_q$ as a fused supervision.

\smallskip \noindent \textbf{Uniform Supervision.}
Another simple strategy is to provide a dense retriever with equal exposures to multiple sources of supervisions. 
Specifically, given a query, we uniformly sample a source of supervision; i.e., a ranking list $\mathcal{P}_q^n$, where $n \sim \mathcal{U}(1,N)$.
This approach naturally encourages the positive samples appearing in more ranking lists to be sampled and vise versa. 
The advantage is that fusion weight tuning is not required.
Furthermore, models can see diverse supervisions from different teachers in contrast to fused supervision, which may be dominated by a single strong teacher (see Appendix~\ref{appedix:intuition} for more study).   

\smallskip \noindent \textbf{Progressive Supervision.} 
The previous two approaches directly give models supervisions from multiple teachers at once; however, learning directly from the mixture of supervision is challenging, especially for DR models which compute text matching with simple dot product. Inspired by the success of curriculum learning~\cite{cldrd}, we propose an approach to \textit{progressive label augmentation} to guide DR training with progressively more challenging supervision. 
Specifically, we train our models with uniform supervision for $N$ iterations and at each iteration, we augment relevance label using additional teacher, as illustrated in Figure~\ref{fig:approach}; i.e., at iteration $T \leq N$, we uniformly sample a source of supervision, $\mathcal{P}_q^n$, where $n \sim \mathcal{U}(1,T)$. 
A key factor of this approach is how to arrange the order for easy-to-hard supervisions; namely, the trajectory of progressive supervision.

With any aforementioned strategy to obtain diverse supervisions, we train our dense retrievers using contrastive loss in Eq.~(\ref{eq:cl}). 
Specifically, given a query $q$, we first obtain a source of supervision either from sampling ($\mathcal{P}_q^n$) or fusion ($\mathcal{F}_q$); then, we randomly sample a positive and hard negative from the top 10 passages and top 46--50 passages, respectively to form a triplet. 
The sampling scheme has been empirically proved to well preserve the supervised signal from a single teacher~ \citep{spar} (also see our study in Appendix~\ref{appedix:ablation_on_positive}). 
In this work, we further extend the sampling scheme to obtain diverse supervisions from multiple teachers.

\subsection{Empirical Studies}
\label{subsec:empirical_study}
\begin{table}[t]
	\caption{Strategies to obtain multiple supervisions using cropped sentences as queries.}
	\label{tb:sampling_ablation}
	\centering
	 \resizebox{0.5\textwidth}{!}{  
     \begin{threeparttable}
    \begin{tabular}{l!{\color{lightgray}\vrule}cccccc}
	\hline \hline
 & 0& 1& 2& 3& 4& 5$\ast$ \\
 \hline
\multirow{2}{*}{Teacher}& \rotatebox[origin=c]{290}{\small uniCOIL}& \rotatebox[origin=c]{290}{\small Contriever}& \rotatebox[origin=c]{290}{\small ColBERTv2}& \multicolumn{3}{c}{three teachers}\\
\cmidrule(lr){5-7}
 & & & & fused& unif.& prog.\\
\arrayrulecolor{black}
\hline
&\multicolumn{6}{c}{effectiveness of student}\\
 \hline
MARCO Dev & 34.9& 33.9& 36.4& 36.7& \textbf{36.9}& 36.6\\
 BEIR-13& 46.7& 47.0& 46.3& 46.6& 47.7& \textbf{49.3}\\
 \hline
&\multicolumn{6}{c}{effectiveness of teacher}\\
 \hline
 MARCO Dev & 35.1& 34.1& 39.7& 40.0& -& -\\
 BEIR-13& -$\tnote{$\bigtriangleup$}$& 47.5& 49.9& -$\tnote{$\bigtriangleup$}$& -& -\\
\arrayrulecolor{black}
	\hline \hline
	\end{tabular}
 \begin{tablenotes}
  \item[$\ast$] The condition of column 5 corresponds to row 0 in Table~\ref{tb:trajectory_ablation}.  
  \item[$\bigtriangleup$] We do not evaluate uniCOIL on BEIR due to its requirement of expensive document expansion from corpus.
    \end{tablenotes}
    \end{threeparttable}
	
 }
\end{table} 
\smallskip \noindent \textbf{Strategies to Obtain Diverse Supervisions.}
We first conduct empirical studies on how to better train a dense retriever in a simplified setting by using the MS MARCO cropped sentences as augmented queries and obtain supervised labels using three teachers with diverse relevance score computation:\ uniCOIL (sparse), Contriever (dense) and ColBERTv2 (multi-vector). 
To compare the different strategies discussed in Section~\ref{subsec:supervsion_strategies}, 
We report the models trained with single (columns 0--2) and multiple (columns 3--5) sources of supervisions for 20 epochs and 60 epochs, respectively. 
For progressive supervision, we follow the supervision trajectory:\ unCOIL$\rightarrow $ Contriever $\rightarrow$ ColBERTv2 with 20 epochs for each of the three iterations ($N=3$). 
Note that we use MS MARCO development queries to tune and obtain the best hyperparameters to create fusion list.   

The results are tabulated in Table~\ref{tb:sampling_ablation}. 
We observe that when learning from a single supervision (columns 0--2), there is a tradeoff between supervised and zero-shot retrieval effectiveness. 
Learning from the fusion list only sees a slight improvement over supervised evaluation while no improvement observes in zero-shot evaluations (columns 0--2 vs 3). 
By contrast, the model sees notable improvements in zero-shot evaluations when trained with uniform supervision (columns 0--3 vs 4), indicating that learning from the diverse relevance labels from multiple retrievers separately rather than single strong supervision (ColBERTv2 or fused supervision) is key to gain generalizability capability. 
Finally, we observe that progressive supervision can further guide a dense retriever to gain generalization capability over uniform supervision (column 4 vs 5). 
Thus, we use progressive supervision in the following experiments.

\begin{table}[t]
	\caption{Study on trajectory of progressive supervision using cropped sentences as queries.}
	\label{tb:trajectory_ablation}
	\centering
	 \resizebox{0.5\textwidth}{!}{  
  \begin{threeparttable}
    \begin{tabular}{l!{\color{lightgray}\vrule}cc}
	\hline \hline
\multicolumn{1}{c!{\color{lightgray}\vrule}}{Progressive supervision}& MARCO dev& BEIR-13 \\
\cmidrule(lr){1-1} \cmidrule(lr){2-2} \cmidrule(lr){3-3} 
\multicolumn{1}{c!{\color{lightgray}\vrule}}{trajectories} & RR@10& nDCG@10\\
\hline
(0) \small unCOIL$\rightarrow $ Contriever $\rightarrow$ ColBERTv2 &36.6& \textbf{49.3}\\
(1) \small Contriever $\rightarrow$ unCOIL$\rightarrow $ ColBERTv2& 36.7& 48.4\\
(2) \small ColBERTv2$\rightarrow $ Contriever$\rightarrow $ unCOIL& 36.4& 47.7\\
(3) \small unCOIL$\rightarrow $ Contriever $\rightarrow$ ColBERTv2$\tnote{$\ast$}$ &\textbf{36.8}& 47.4\\
	\hline \hline
	\end{tabular}
 \begin{tablenotes}
	\item[$\ast$] ColBERTv2 is the only teacher at the last (3rd) iteration. 
    \end{tablenotes}
    \end{threeparttable}
	}
\end{table} 
\smallskip \noindent \textbf{Trajectory of Progressive Supervision.}
We then study how to better arrange the trajectories of progressive supervision in Table~\ref{tb:trajectory_ablation}. 
We observe that different trajectories have much impact on models' zero-shot retrieval effectiveness while a minor impact on supervised evaluation can be seen. 
For example, switching the sampling order between uniCOIL and Contriever results in a degrade of 1 point on the averaged nDCG@10 over BEIR-13 (column 0 vs 1) while reversing the whole trajectory leads to a degrade with more than 1.5 points (column 0 vs 3).  
This observation reflects an intuition that the retrievers with better generalization capability may capture more complex matching signal between text pairs (ColBERTv2 shows better generalization capability than the other two teachers); thus, their relevance labels should be augmented at a later stage of model training. 
Finally, in column 3, we follow the trajectory in column 0 but only use ColBERTv2 as the only source of supervision instead of obtaining uniform supervision from the three teachers at the last iteration. 
This change results in worse zero-shot retrieval effectiveness, indicating that learning from diverse supervisions is key to training a generalizable dense retriever.

\begin{figure}[t]
\begin{subfigure}{0.49\columnwidth}
\includegraphics[width=\columnwidth]{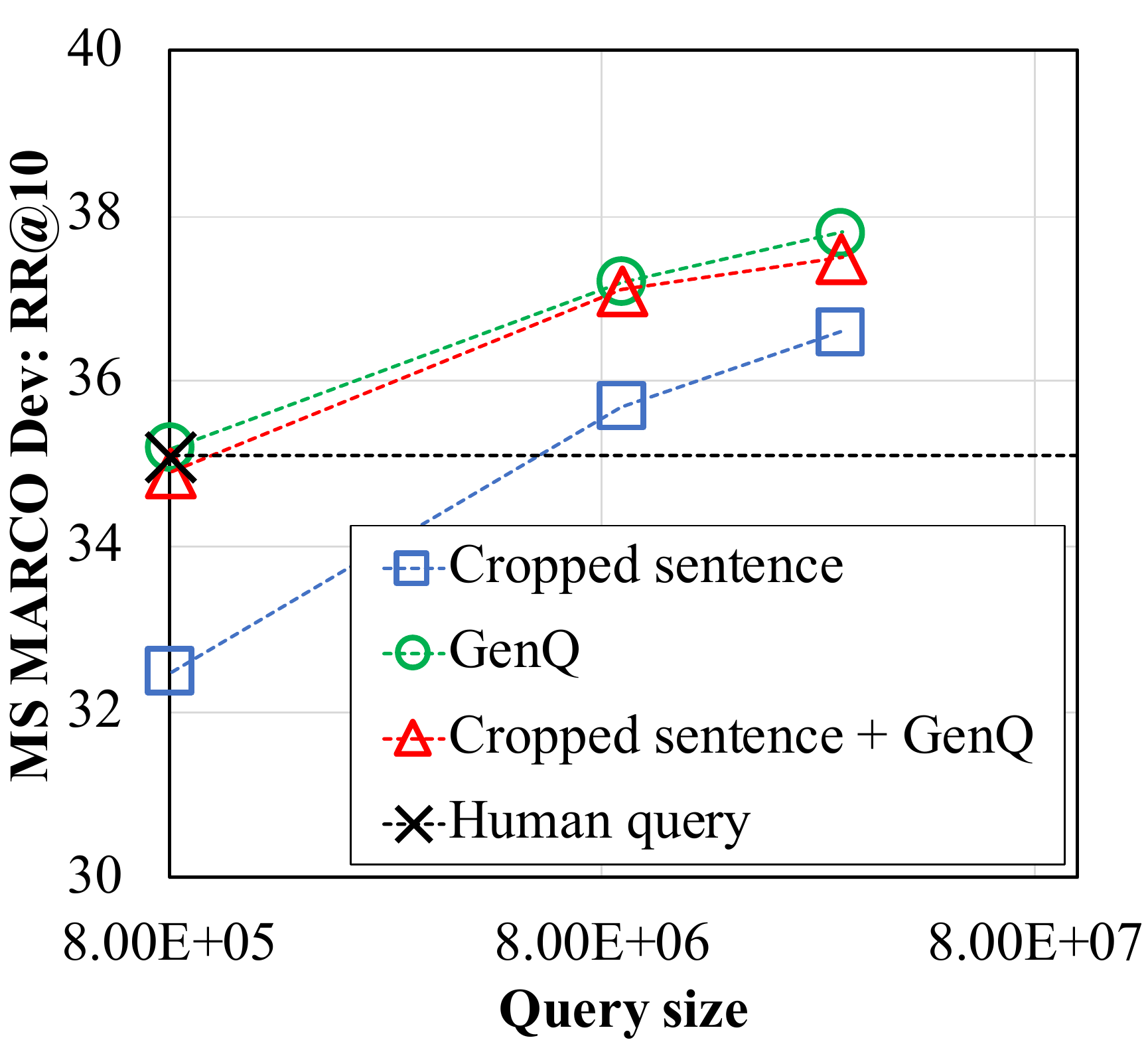}
\caption{MS MARCO Dev}
\end{subfigure}
\begin{subfigure}{0.49\columnwidth}
\includegraphics[width=\columnwidth]{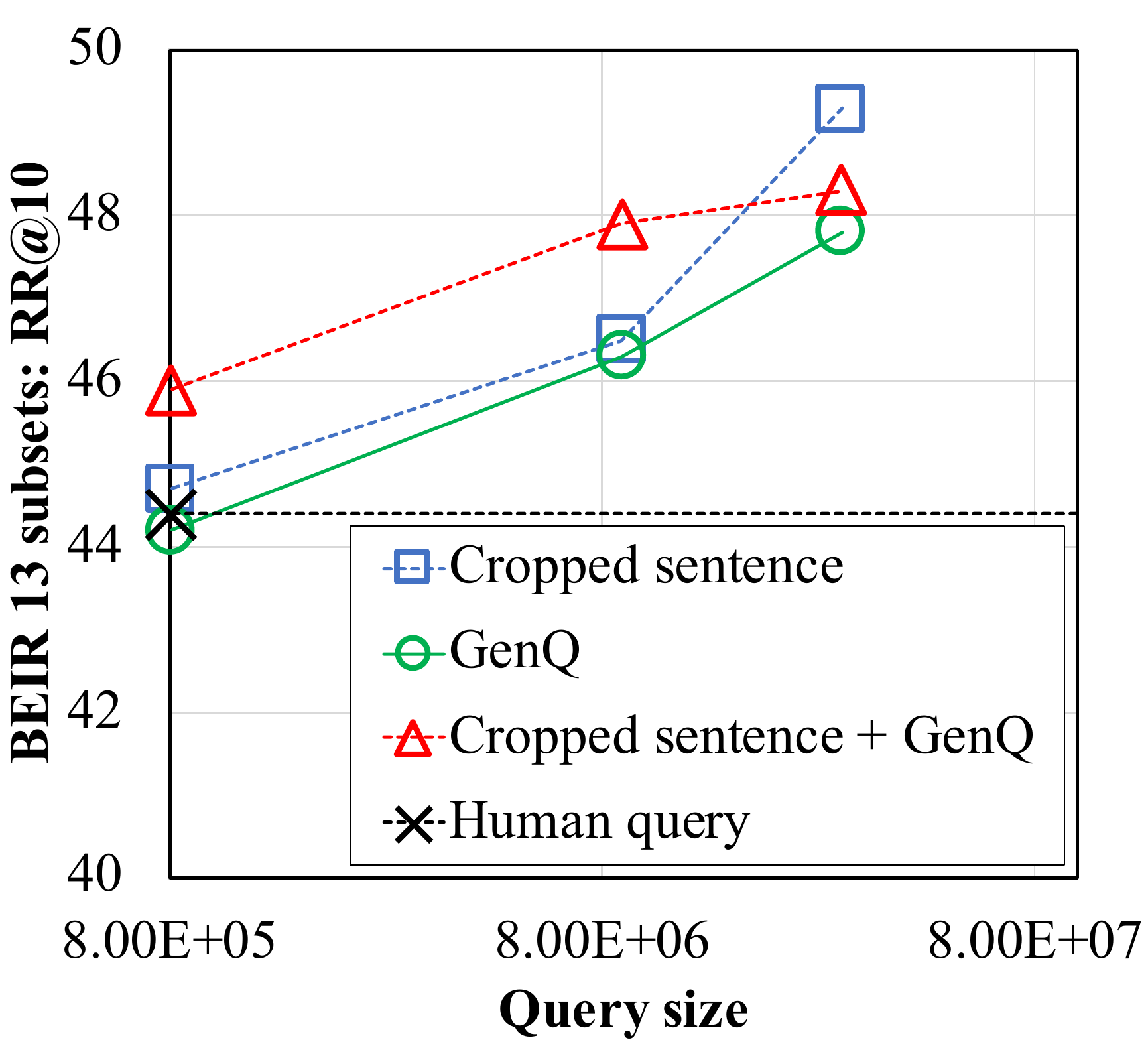}
\caption{BEIR-13}
\end{subfigure}
\caption{Impacts of query augmentation.}
\label{fig:query_augmentation}
\end{figure}

\smallskip \noindent \textbf{Query Augmentation.}
 We study the impacts of query augmentation on models' effectiveness under the scenario where supervision is given, which has not been studied so far in the field of dense retrieval.  
 With the best trajectory of progressive supervision, we compare models trained using cropped sentences (rectangles), generative queries (GenQ; circles), their mixture (triangles) and human queries (cross) in Figure~\ref{fig:query_augmentation}. 
We observe that query size is the key to successful training. 
Although training with limited (0.8M) cropped sentences cannot perform well in MS MARCO dataset, scaling up the size (28M) sees significant improvement over the model trained on 0.8M human queries. 
Similarly, in Figure~\ref{fig:query_augmentation}~(b), model's generalization capability shows a huge jump when scaling up query size and surprisingly, cropped sentences can help dense retrievers to gain more generalization capability than human-like generative queries. 
However, a mixture of cropped sentences and generative queries yields strong retrieval effectiveness in supervised and zero-shot evaluations, especially when the query size is not large enough (0.8--8M).

\subsection{Training our \ourmodel s} 
With the empirical studies on DR training, we then propose the final recipe to train our \ourmodel.  
We train \ourmodel\ for 20 epochs (around 130K steps) at each iteration, with the trajectory of progressive supervision:\ unCOIL$\rightarrow $ Contriever $\rightarrow$ GTR-XXL $\rightarrow$ ColBERTv2 $\rightarrow$ SPLADE++. 
We list all the teacher model checkpoints for label augmentation in Appendix~\ref{appedix:model_checkpoints}. 
This trajectory is based on models' retrieval effectiveness on BEIR with the intuition gained from Section~\ref{subsec:empirical_study} that a more generalizable model creates more complex relevance labels.  
For query augmentation, we mix half of cropped sentences and synthetic queries as training queries. 
Note that we do not further fine-tune our models on the MS MARCO training set.
In addition, we train other three \ourmodel\ variants. 
\ourmodel-S and \ourmodel-Q only use cropped sentences and synthetic queries, respectively. 
As for \ourmodel+, we follow the same training procedure of \ourmodel\ but switch the initialization from BERT to the masked auto-encoding pre-trained model, RetroMAE\footnote{\url{https://huggingface.co/Shitao/RetroMAE}}. 
We will discuss the impacts of initialization in Section~\ref{sec:discussion}. 
The implementation of \ourmodel s and the fully augmented training data using the five teachers are detailed in Appendix~\ref{appedix:implementation_details} and \ref{appedix:data_statistics}, respectively.

\section{Comparison with the State of The Art}\label{sec:experiments}
\subsection{Datasets}
\label{subsec:datasets}
In addition to MS MARCO dev, we evaluate model supervised effectiveness on the TREC DL~\cite{trec19dl, trec20dl} queries, created by the organizers of the 2019 (2020) Deep Learning Tracks at the Text REtrieval Conferences (TRECs), where 43 (53) queries with on average 95 (68) graded relevance labels per query (in contrast to 6980 queries with on average 1 non-graded relevance label per query in MS MARCO dev) are released. 
We report nDCG@10, used by the organizers as the main metric. 

For zero-shot evaluations, we evaluate models on all the 18 datasets in  BEIR~\cite{beir}. 
In addition, we use LoTTE~\cite{colbert-v2} consisting of questions and answers posted on StackExchange with five topics including writing, recreation, science,
technology, and lifestyle. 
We evaluate models' retrieval effectiveness in the pooled setting, where the passages and
queries from the five topics are aggregated. 
Following \citet{colbert-v2}, the retrieval effectiveness of Success@5 on search and forum queries are reported. 
We refer reader to Appendix~\ref{appedix:lotte} for detailed evaluations on all the five topics.

\subsection{Baseline Models}
We compare \ourmodel~with dense retrievers using the backbone of bert-base-uncased trained with advanced techniques.
(1) Knowledge Distillation:\ RocketQAv2~\cite{rocketqav2} distills knowledge from a cross encoder while CL-DRD~\cite{rocketqav2} combines curriculum learning and cross-encoder distillation. 
They all use cross encoders' knowledge to augment positive relevance lablels as our approach. 
(2) Contrastive Pre-Training:\ coCondneser~\cite{cocondenser}, Contriever~\cite{contriever} and COCO-DR~\cite{cocodr} are first pre-trained on different corpus listed in Table~\ref{tb:model_comparison}, and then fine-tuned on MS MARCO training queries. 
(3) Masked Auto-Encoding Pre-Training:\ COT-MAE~\cite{cotmae} and RetroMAE~\cite{retromae} are first pre-trained to recover polluted sentences and then fine-tuned on MS MARCO training queries. For RetroMAE, we use the variant with the best BEIR retrieval effectiveness for comparison. 
(4) Domain adaptation: \ We consider GPL~\cite{gpl} and Promptgator~\citep[PTR;][]{promptgator}, which use generative models to create pseudo relevance data for each corpus in BEIR and train one expert dense retriever for each corpus. 
This approach requires the target corpus while training. 
Note that COCO-DR can also be considered domain adaptation on BEIR although it uses one model for all tasks. 

Note that coCondenser and COT-MAE are fine-tuned on the “non-standard” MS MARCO passage corpus that has been augmented with title. 
Thus, we also conduct inference on the corpus with title for them; otherwise, we use the official MS MARCO passage corpus. 
In addition, we also report the retrieval effectiveness of GTR-XXL~\cite{gtr} and ColBERTv2~\cite{colbert-v2} from their original papers and conduct retrieval for SPLADE++ using pyserini~\cite{pyserini} for reference.
We list all the other model checkpoints used for evaluations in Appendix~\ref{appedix:model_checkpoints}.

\subsection{Implementation Details}
\label{appedix:implementation_details}
We train our dense retrievers initialized from bert-base-uncased on 32 A100 GPUs with a per-GPU batch size of 64 and a learning rate of
$3e-5$.
Each batch includes an augmented query with its positives and hard negatives. 
Following \citet{dpr}, we use asymmetric dual encoder with two distinctly parameterized encoders and leverage in-batch negative mining. 
Note that symmetric dual encoder shows poor generalization capability in our initial experiments. 
We set the maximum query and passage lengths to 32 and 128 for MS MARCO training and evaluation. 
For BEIR evaluation, we set maximum input lengths to 512.

\subsection{Results}
\label{subsec:comparison_to_sota}
\smallskip \noindent \textbf{Supervised Evaluations.}
The first main row in Table~\ref{tb:main_comparison} reports models' retrieval effectiveness on MS MARCO passage ranking dataset. 
We first observe that some baseline dense retrievers which perform well in MS MARCO dev set are either pre-trained on MS MARCO corpus (coCondenser and COT-MAE) or well fine-tuned on MS MARCO training queries with cross-encoder distillation (CL-DRD and RocketQAv2). 
However, their retrieval effectiveness on MS MARCO dev set is not well correlated to TREC DL queries, which have fine-grained human labels with different degrees of relevance. 
We hypothesize that these models are able to retrieve the most relevant passage from the corpus but cannot retrieve diverse passages with different degrees of relevance. 
By contrast, all the variants of \ourmodel\ trained with diverse augmented relevance labels show consistenly strong retrieval effectiveness in MS MARCO dev and TREC DL queries, over RR@10 38.0 and nDCG@10 70.0, respectively. 

\begin{table*}[t]
	\caption{Comparison with existing state-of-the-art dense retrievers. Bold (underline) denotes the best (second best) effectiveness for each row among baseline dense models.}
	\label{tb:main_comparison}
	\centering
	 \resizebox{1\textwidth}{!}{  
	 \begin{threeparttable}
	  	 \setlength\tabcolsep{4pt}
    \begin{tabular}{l!{\color{lightgray}\vrule}ccc!{\color{lightgray}\vrule}ccccccccc!{\color{lightgray}\vrule}cccc}
	\hline \hline

Rep type&sparse& mul-vec& \multicolumn{1}{c!{\color{lightgray}\vrule}}{dense}& \multicolumn{9}{c!{\color{lightgray}\vrule}}{dense (baselines)}& \multicolumn{4}{c}{dense (ours)}\\ \hline
& 0& 1& 2& 3& 4& 5& 6& 7& 8& 9& A& B& C& D& E& F\\
\hline
&\rotatebox[origin=c]{290}{SPLADE\small++}&
\rotatebox[origin=c]{290}{ColBERT\small v2}&
\rotatebox[origin=c]{290}{GTR-XXL}&
\rotatebox[origin=c]{290}{CL-DRD}&
\rotatebox[origin=c]{290}{RocketQAv2}&
 \rotatebox[origin=c]{290}{COT-MAE}&
 \rotatebox[origin=c]{290}{RetroMAE}&
\rotatebox[origin=c]{290}{coCondenser}&
\rotatebox[origin=c]{290}{Contriever}&
 \rotatebox[origin=c]{290}{COCO-DR}&
\rotatebox[origin=c]{290}{GPL}&
 \rotatebox[origin=c]{290}{PTR}&
 \rotatebox[origin=c]{290}{\ourmodel-S}&
 \rotatebox[origin=c]{290}{\ourmodel-Q}&
 \rotatebox[origin=c]{290}{\ourmodel}&
 \rotatebox[origin=c]{290}{\ourmodel+}\\
\hline
  Pre-training &\cmark  & \xmark &  \cmark& \xmark & \xmark & \cmark & \cmark& \cmark& \cmark& \cmark& \xmark& \cmark& \xmark& \xmark& \xmark& \cmark\\
  \arrayrulecolor{lightgray}
  \hline
 Distillation&  \cmark  & \cmark & \cmark &  \cmark &  \cmark &  \xmark& \xmark& \xmark& \xmark& \xmark& \cmark& \xmark& \cmark& \cmark& \cmark& \cmark\\
 \hline
 Target Corpus$\tnote{$\dagger$}$&  \xmark  & \xmark& \xmark &  \xmark &  \xmark &  \xmark& \xmark& \xmark& \xmark& \cmark& \cmark& \cmark& \xmark& \xmark& \xmark& \xmark\\
 \arrayrulecolor{black}
\hline
nDCG@10&\multicolumn{16}{c}{\textbf{MS MARCO (Supervised)}}\\
 \arrayrulecolor{lightgray}
\hline
Dev (RR@10)& 38.9& 39.7& 38.8& 38.1& 38.8$\tnote{$\ast$}$& \textbf{39.9}$\tnote{$\ast$}$& 35.4& 38.6$\tnote{$\ast$}$& 34.1& 35.8& -& -& 38.1& 39.1& \underline{39.3}& 39.0\\
Dev (R@1K)& 98.2& 98.4& 99.0& 97.9& 98.1$\tnote{$\ast$}$& 98.5$\tnote{$\ast$}$& 97.5& 98.4$\tnote{$\ast$}$& 97.9& 97.9& -& -& 98.3& \textbf{98.8}& 98.5& \underline{98.6}\\
DL2019& 74.3& 74.6& -& 72.5& -& 70.0$\tnote{$\ast$}$& 68.8& 71.5$\tnote{$\ast$}$& 67.8& \underline{74.1}& -& -& 73.6& 74.0& 74.1& \textbf{74.4}\\
DL2020& 71.8& 75.2& -& 68.3& -& 67.8$\tnote{$\ast$}$& 71.4& 68.1$\tnote{$\ast$}$& 66.1& 69.7& -& -& 70.0& \underline{72.6}& \textbf{72.9}& 72.3\\
\arrayrulecolor{black}
\hline
nDCG@10& \multicolumn{16}{c}{\textbf{BEIR (Zero-shot)}}\\
\arrayrulecolor{lightgray}
\hline
TREC-COVID& 71.1& 73.8& 50.1& 58.4& 67.5& 56.1& \underline{77.2}& 71.2& 59.6& \textbf{78.9}& 70.0& 72.7& 73.9& 73.2& 74.0& 75.9\\
NFCorpus& 34.5& 33.8& 34.2& 31.5& 29.3& 32.1& 30.8& 32.5& 32.8& \textbf{35.5}& \underline{34.5}& 33.4& 32.2& 33.0& 32.9& 33.9\\
FiQA-2018& 35.1& 35.6& 46.7& 30.8& 30.2& 28.3& 31.6& 27.6& 32.9& 31.7& 34.4& \textbf{40.4}& \underline{35.6}& 35.3& 35.0& \underline{35.6}\\
ArguAna& 52.1& 46.3& 54.0& 41.3& 45.1& 27.8& 43.3& 29.9& 44.6& 49.3& \textbf{55.7}& \underline{53.8}& 51.5& 45.5& 48.9& 46.9\\
Tóuche-2020& 24.4& 26.3& 25.6& 20.3& 24.7& 21.9& 23.7& 19.1& 23.0& 23.8& 25.5& \textbf{26.6}& \underline{26.5}& 26.0& 24.9& 26.3\\
Quora& 81.4& 85.2& 89.2& 82.6& 74.9& 75.6& 84.7& 85.6& 86.5& 86.7& 83.6& -& 86.4& \underline{87.1}& 86.9& \textbf{87.5}\\
SCIDOCS& 15.9& 15.4& 16.1& 14.6& 13.1& 13.2& 15.0& 13.7& \underline{16.5}& 16.0& \textbf{16.9}& 16.3& 15.9& 15.0& 15.4& 15.9\\
SciFact& 69.9& 69.3& 66.2& 62.1& 56.8& 60.1& 65.3& 61.5& 67.7& \textbf{70.9}& 67.4& 62.3& 67.8& 67.2& 67.5& \underline{67.9}\\
NQ& 54.4& 56.2& 56.8& 50.0& 50.5& 48.3& 51.8& 48.7& 49.5& 50.5& 48.3& -& \underline{53.3}& 52.3& 53.1& \textbf{53.7}\\
HotpotQA& 68.6& 66.7& 59.9& 58.9& 53.3& 53.6& 63.5& 56.3& 63.8& 61.6& 58.2& 60.4& \underline{65.6}& 62.7& 64.8& \textbf{66.2}\\
DBPedia& 44.2& 44.6& 40.8& 38.1& 35.6& 35.7& 39.0& 36.3& 41.3& 39.1& 38.4& 36.4& 40.6& 41.4& \underline{41.4}& \textbf{41.7}\\
FEVER& 79.6& 78.5& 74.0& 73.4& 67.6& 50.6& 77.4& 49.5& 75.8& 75.1& 75.9& 76.2& \underline{76.4}& 75.1& 75.8& \textbf{78.1}\\
Climate-FEVER& 22.8& 17.6& 26.7& 20.4& 18.0& 14.0& 23.2& 14.4& \textbf{23.7}& 21.1& \underline{23.5}& 21.4& 21.8& 20.3& 22.2& 22.7\\
CQADupStack& 34.1& -& 39.9& 32.5& -& 29.7& 34.7& 32.0& 34.5& \textbf{37.0}& 35.7& -& \underline{35.9}& 34.4& 35.2& 35.4\\
Robust04& 45.8& -& 50.6& 37.7& -& 30.8& 44.7& 35.4& \underline{47.6}& 44.3& 43.7& -& 46.3& 45.3& 47.2& \textbf{47.9}\\
Signal-1M& 29.6& -& 27.3& 28.2& -& 21.1& 26.5& 28.1& 19.9& 27.1& 27.6& -& \underline{29.3}& 28.9& \textbf{30.1}& \underline{30.1}\\
TREC-NEWS& 39.4& -& 34.6& 38.0& -& 26.1& \underline{42.8}& 33.7& \underline{42.8}& 40.3& 42.1& -& 42.1& 40.0& 41.6& \textbf{44.4}\\
BioASQ& 50.4& -& 32.4& 37.4& -& 26.2& 42.1& 25.7& 38.3& 42.9& \textbf{44.2}& -& 41.3& 41.9& 41.7& \underline{43.3}\\
\arrayrulecolor{lightgray}
\hline
\multicolumn{2}{l}{Averaged nDCG@10} \\
\hline
PTR 11 subsets& 47.1& 46.2& 44.9& 40.9& 40.1& 35.7& 44.5& 37.4& 43.8& 45.7& 45.5& 45.5& \underline{46.2}& 45.0& 45.7& \textbf{46.5}\\
BEIR-13& 50.3& 49.9& 49.3& 44.8& 43.6& 39.8& 48.2& 42.0& 47.5& 49.2& 48.6& -& \underline{49.8}& 48.8& 49.4& \textbf{50.2}\\
All& 47.4& -& 45.8& 42.0& -& 36.2& 45.4& 38.9& 44.5& 46.2& 45.9& -& \underline{46.8}& 45.8& 46.6& \textbf{47.4}\\
\arrayrulecolor{black}
\hline
Success@5& \multicolumn{16}{c}{\textbf{LoTTE (Zero-shot)}}\\
\arrayrulecolor{lightgray}
\hline
Search (pooled)& 70.9& 71.6& -& 65.8& 69.8& 63.4& 66.8& 62.5& 66.1& 67.5& -& -& 71.4& 72.4& \underline{72.6}& \textbf{73.5}\\
Forum (pooled)& 62.3& 63.4& -& 55.0& 57.7& 51.9& 58.5& 52.1& 58.9& 56.8& -& -& 61.1& 61.2& \underline{61.4}& \textbf{62.1}\\
\arrayrulecolor{black}
	\hline \hline
	\end{tabular}
		\begin{tablenotes}
	\item[$\dagger$] The approach assumes the target corpus (e.g., BEIR) is available while training. 
    \item[$\ast$] These numbers are not comparable due to the use of a ``non-standard'' MS MARCO passage corpus augmented with title. 
    \end{tablenotes}
    \end{threeparttable}
	}
\end{table*} 
\smallskip \noindent \textbf{Zero-Shot Evaluations.} 
The second main row in Table~\ref{tb:main_comparison} reports models' zero-shot retrieval effectiveness on the BEIR datasets. 
We observe a reverse trend that those dense retrievers performing relatively poorly in MS MARCO dev queries have better zero-shot retrieval effectiveness (e.g., Contriever, COCO-DR and RetroMaE). 
These models are pre-trained (with data augmentation) on a corpus other than MS MARCO to combat domain shift issue in dense retrieval~\cite{modir, cocodr}. 
On the other hand, our models trained on augmented data from MS MARCO corpus only transfer well to BEIR datasets. 
For example, \ourmodel+ reaches state-of-the-art retrieval effectiveness on BEIR as the sparse retriever, SPLADE++.\footnote{See our BEIR leaderboard submission \href{https://eval.ai/web/challenges/challenge-page/1897/leaderboard/4475}{here}.}
In addition, all the \ourmodel\ variants outperform dense retrievers by a large margin and compete SPLADE++ and ColBERTv2 in the LoTTE dataset. 
It is worth mentioning that the models trained with domain adaptation (columns 9--B) perform better than the others (columns 3--8) but still underperform \ourmodel\ in zero-shot evaluations. 
Using \ourmodel\ as a foundation model for domain adaptation is possible to gain DR zero-shot effectiveness, which we leave to our future work. 

To sum up, \ourmodel s advances state-of-the-art zero-shot effectiveness on BEIR and LoTTE while keeping strong effectiveness on supervised evaluation on MS MARCO. 
The experimental results demonstrate that our data augmentation approaches enable dense retrievers to learn domain-invariant matching signal between text pairs as the other models with fine-grained late interaction (SPLADE++ and ColBERTv2) or 40 times larger model size (GTR-XXL). 

A comparison between \ourmodel-S and \ourmodel-Q (columns C and D) shows that augmented query type has an impact on retrieval effectiveness in different datasets. 
\ourmodel-S trained on cropped sentences surprisingly sees the highest retrieval effectiveness on BEIR but only sacrifices a bit on MS MARCO datasets. 
This means that cropped sentences, the cheap query type (compared to neural generative queries), are sufficiently helpful for models to learn domain-invariant retrieval capability. 
By contrast, we observe that \ourmodel-Q trained with human-like queries performs poorly compared to \ourmodel-S on the datasets where queries are far different from human-like queries, such as ArguAna (45.5 vs 51.5) and CQADupStack (34.4 vs 35.9), while mixing different types of queries (\ourmodel) can mitigate the issue. 
Finally, \ourmodel+ combined with masking auto-encoding (MAE) pre-training and our approach sees further improvement on zero-shot evaluations without sacrificing in-domain ones, indicating that MAE pre-training may be orthogonal to our approach based on contrastive learning.

\section{Discussions} 
\label{sec:discussion}
\smallskip \noindent \textbf{Is it necessary to augment relevance labels with a cross encoder?} 
To answer this question, we further train DRAGON-S with the augmented relevance labels from a cross encoder (CE). 
Specifically, we create a rank list of CE by first retrieving top 1000 passages with DRAGON-S and re-rank them with the CE for each cropped sentence. 
With the CE ranking list, we conduct another iteration (20 epochs) of training for DRAGON-S; however, we do not see retrieval effectiveness improvement in Table~\ref{tb:ce_ablation} (column 0 vs 1). 
In addition, the retrieval effectiveness becomes even worse when we further train DRAGON-S by only sampling CE ranking list instead of uniformly sampling all the six ranking lists (column 1 vs 2). 
Finally, we initialize from bert-base-uncased and re-train the model for three iterations (60 epochs) only with the CE ranking list.\footnote{We do not notice effective improvement with more iterations of training both in supervised and zero-shot evaluations.} 
We observe that its effectiveness (column 3) is even worse than the models trained with the ranking lists from three retrievers (see columns 4 and 5 in Table~\ref{tb:sampling_ablation}). 
This result contradicts the general belief that CE provides the strongest supervision to a dense retriever. 
Moreover,  it demonstrates the effectiveness of using diverse supervisions to train a generalizable dense retriever, rather than relying on a single strong supervision. 

Table~\ref{tb:rla_cost} compares the latency cost per query for relevance label augmentation with different neural rankers and demonstrates that leveraging all the retrievers to augment relevance labels are still more efficient than a cross encoder. 
We detail the measurement setting in Appendix~\ref{appedix:latency_measurement_for_rla}.

\begin{table}[t]
	\caption{Label augmentation with CE (miniLML6v2) using cropped sentences as queries. All denotes the five sources of supervisions used for training \ourmodel-S.}
	\label{tb:ce_ablation}
	\centering
	 \resizebox{0.5\textwidth}{!}{  
  \begin{threeparttable}
    \begin{tabular}{l!{\color{lightgray}\vrule}cccc}
	\hline \hline
& 0$\tnote{$\ast$}$& 1& 2& 3\\
\hline
\ourmodel-S initialization& & \checkmark& \checkmark& \\
Source of supervision& & all + CE & CE only& CE only\\
\hline
MARCO Dev (RR@10)& \textbf{38.1}& \textbf{38.1}& 37.5& 36.8\\
BEIR-13 (nDCG@10)& \textbf{49.8}& 49.7& 48.7& 47.7\\
\arrayrulecolor{black}
	\hline \hline
	\end{tabular}
 \begin{tablenotes}
	\item[$\ast$] Column 0 corresponds to \ourmodel-S. 
    \end{tablenotes}
    \end{threeparttable}
	}
\end{table} 
\begin{table}[t]
	\caption{The latency comparison of relevance label augmentation with batch inference using different teachers on MS MARCO.}
	\label{tb:rla_cost}
	\centering
	 \resizebox{0.5\textwidth}{!}{
    \begin{tabular}{llr!{\color{lightgray}\vrule}rr}
	\hline \hline
 & & candidates & \multicolumn{2}{c}{latency (ms/q)} \\
 \cmidrule(lr){3-3} \cmidrule(lr){4-5} 
 \multicolumn{1}{l}{Type}& \multicolumn{1}{l}{Model}& (\#)& \multicolumn{1}{r}{GPU}& \multicolumn{1}{r}{CPU}\\ 
\hline
\arrayrulecolor{lightgray}
cross-encoder &miniLML6v2& 1K& 600& -\\
dense &Contriever& 8.8M& < 1& -\\
dense &GTR-XXL& 8.8M& 10& -\\
sparse &uniCOIL& 8.8M& -& 84\\
sparse &SPLADE++& 8.8M& -& 144\\
multi-vec &ColBERTv2& 8.8M& 55& -\\
\arrayrulecolor{black}
	\hline \hline
	\end{tabular}
	}
\end{table} 
\begin{table}[t]
	\caption{Ablation on initialized checkpoint using the mixture of cropped sentences and GenQ as queries.}
	\label{tb:initialization_ablation}
	\centering
	 \resizebox{0.42\textwidth}{!}{  
  \begin{threeparttable}
    \begin{tabular}{lcc}
	\hline \hline
\multirow{2}{*}{Initialized checkpoint}& MARCO dev& BEIR-13 \\
\cmidrule(lr){2-2} \cmidrule(lr){3-3} 
& RR@10& nDCG@10\\
\hline
(0) BERT base &\textbf{39.3}& 49.4\\
(1) Contriever& 38.7& 49.3\\
(2) RetroMAE& 39.0& \textbf{50.2}\\
	\hline \hline
	\end{tabular}
 \begin{tablenotes}
	\item[$\ast$] Row 0 and 2 corresponds to  \ourmodel \ and \ourmodel+, respectively. 
    \end{tablenotes}
    \end{threeparttable}
	}
\end{table} 

\begin{figure*}[t]
    \centering
    \resizebox{1\textwidth}{!}{
        \includegraphics{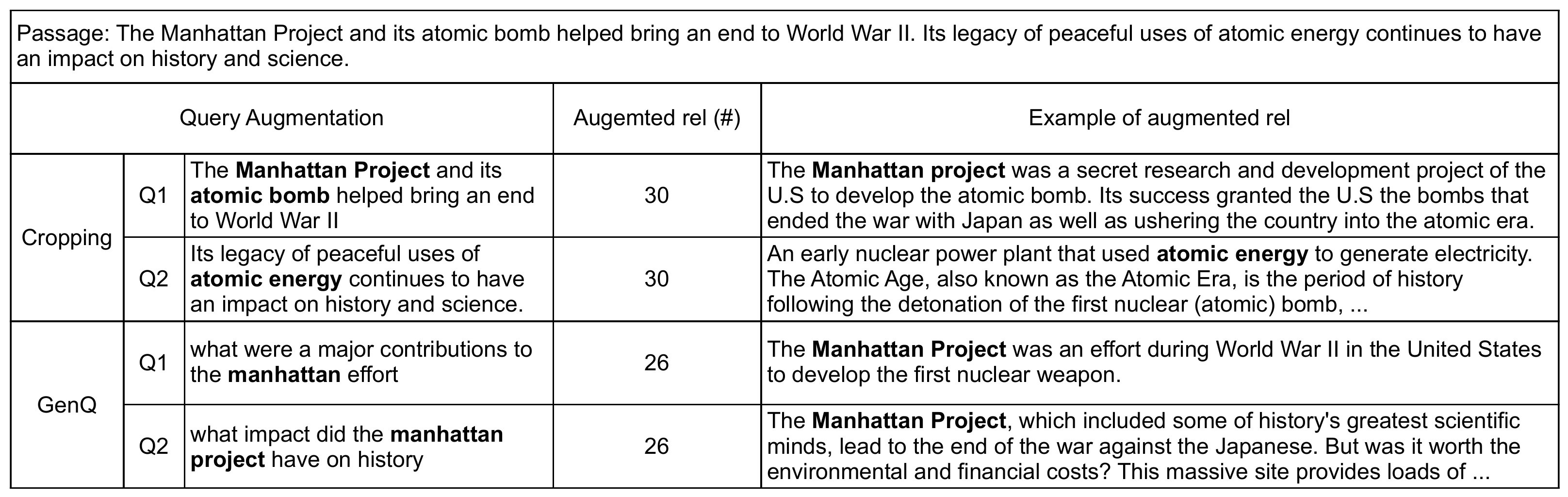}
    }
    \caption{Examples of augmented queries and relevance labels from a passage. Augmented rel (\#) denotes the number of unique relevant passages labeled by all our five teachers.} 
    \label{fig:augmentation_example}
\end{figure*}

\smallskip \noindent \textbf{Does \ourmodel \ benefit from unsupervisd pre-training?} 
Table~\ref{tb:initialization_ablation} compares the models trained from different checkpoint initialization. 
Note that for Contriever and RetroMAE, we initialize from the checkpoint with only unsupervised pre-training (without fine-tuning on MS MARCO). 
We observe that our approach benefits from masked auto-encoding rather than contrastive pre-training (row 2 vs 1). 
The result is sensible since our approach can be considered an improved version of contrastive pre-training, which appears to be orthogonal to masked auto-encoding pre-training. 
We leave the investigation of improving DR with generative and contrastive pre-training combined for future research.

\smallskip \noindent \textbf{Can we use the soft labels from multiple teachers?} 
In the literature, using the relevance scores from a teacher as soft labels is a standard of knowledge distillation~\cite{tct, tasb}. 
However, in our study, even when training with uniform supervision from a sparse and dense retriever (i.e., uniCOIL and Contriever), it is challenging to normalize their relevance scores and create universal soft labels, yielding significant supervised and zero-shot effectiveness drops. 
We suspect that dense and sparse retrievers have many different views on relevance score computation; thus, it is even harder for a dense retriever to learn the score distributions from the different teachers.

\smallskip \noindent \textbf{Why sentence cropping yields a generalizable dense retriever?} 
Figure~\ref{fig:augmentation_example} showcases the augmented queries by sentence cropping and neural generation and their respectively augmented relevant passages other than the original passages. 
We observe two main differences between the cropped sentences and generative queries. 
Cropped sentences provide diverse queries from a passage; i.e., the two cropped sentences in Figure~\ref{fig:augmentation_example} include slightly different topics (Manhattan Project and atomic energy). 
By contrast, all generative queries surround the same main topic, Manhattan Project, about the original passages. 
Second, the cropped sentences have more unique augmented relevant passages than generative queries. 
This is maybe because a cropped sentence, containing more information (keywords), is more challenging than a generative human-like query. 
Thus, teachers show more disagreement between each other on cropped sentences. 
We hypothesize that a dense retriever trained on cropped sentences can capture more diverse supervised signals from multiple teachers than generative queries. 
This explains the reason why \ourmodel-S shows better generalization capability than \ourmodel-Q.

\section{Related Work}
\smallskip \noindent \textbf{Knowledge Distillation.} 
Our work is closely related to the previous work exploring knowledge distillation~\citep[KD;][]{kd} from ColBERT, cross encoder or their ensemble~\cite{tasb, mse_margin} to improve the effectiveness of DR~\cite{tct,rocketqa}. 
However, they only take the advantage of soft labels from KD and use the relevant passages labeled by humans. 
The recent work~\cite{rocketqav2, cldrd} mines more positive samples using cross encoder to further augment the limited relevance labels by humans. 
Nevertheless, it is challenging for cross encoders to augment relevance labels for queries in scale due to its low efficiency. 
\citet{spar} first explore label augmentation using singe sparse retrieval model on large-scale queries and demonstrate that a dense retriever can mimic a teacher of a sparse retriever (e.g., BM25). 
Different from the previous work, we explore label augmentation using multiple supervisions on large-scale augmented queries. 

\smallskip \noindent \textbf{Curriculum Learning.} 
Easy-to-hard training strategies~\cite{cl} have been applied to improve many machine learning tasks, including dense retrieval~\cite{cldrd, prod}. 
The previous work focuses on distilling complex knowledge from cross encoders to a dense retriever with a curriculum training strategy and demonstrates improved effectiveness in supervised retrieval tasks. 
In our work, we explore to progressively train a dense retriever with the diverse supervisions from dense, sparse and multi-vector retrievers to improve both supervised and zero-shot effectiveness.

\smallskip \noindent \textbf{Pre-training.}
There are two popular approaches to pre-training a dense retriever. 
The first one is contrastive pre-training, aiming to increase the size of training data by creating artificial text pairs~\cite{ict, Chang2020Pre-training, contriever} from a corpus or collecting question--answer pairs~\cite{domain-match-pretraining, gtr} from websites. 
The second one is masked auto encoding pre-training, where models are trained to recover the corrupted texts~\cite{condenser, seed-encoder, retromae, cotmae}. 
Our work is similar to contrasitive pre-training but instead of creating large-scale training data in an unsupervised or weakly supervised manner, we investigate how to conduct supervised contrastive learning on artificially created text pairs. 
We demonstrate that combining masked auto encoding pre-training and our supervised contrastive learning can further improve models' generalization capability.

\section{Conclusion} 
We present \ourmodel, \textbf{D}ense \textbf{R}etriever trained with diverse \textbf{A}u\textbf{G}mentati\textbf{ON}. 
We propose a unified framework of data augmentation (DA) to understand the recent progress of training dense retrievers. 
Based on the framework, we extensively study how to improve dense retrieval training through query and relevance label augmentation. 
Our experiments uncover some insights into training a dense retriever, which contradicts common wisdom that cross encoder is the most effective teacher and human-like queries are the most suitable training data for dense retrieval.  
Instead, we propose a diverse data augmentation recipe, query augmentation with the mixture of sentence cropping and generative queries, and progressive relevance label augmentation with multiple teachers. 
With our recipe of DA, we are the first to demonstrate that a single BERT-base-sized dense retriever can achieve state-of-the-art effectiveness in both supervised and zero-shot retrieval tasks. 
We believe that \ourmodel\ can serve as a strong foundation retrieval model for domain adaptation retrieval tasks~\cite{gpl, promptgator} or the existing retrieval augmented language models~\cite{izacard_few-shot_2022, blackboxlm, memorizelm}.

% \section*{Acknowledgements} 

% \newpage

\bibliographystyle{acl_natbib}
\bibliography{paper.bib}

\clearpage

\appendix
\section{Appendices}
\subsection{Impacts of Top-$k$ Positive Sampling}
\label{appedix:ablation_on_positive}
\begin{table}[h]
	\caption{Ablation on progressive label augmentation from top-$k$ passages using cropped sentences as queries.}
	
 \label{tb:positive_ablation}
	\centering
	 \resizebox{0.4\textwidth}{!}{  
  \begin{threeparttable}
    \begin{tabular}{l!{\color{lightgray}\vrule}cccc}
	\hline \hline
top-$k$ positives& 1 & 5& 10\\
\hline
MARCO Dev (RR@10)& 33.1& 36.4& 36.6\\
BEIR-13 (nDCG@10)& 42.4& 48.0& 49.3\\
\arrayrulecolor{black}
	\hline \hline
	\end{tabular}
 \begin{tablenotes}
	 \item[$\ast$] Trajectory: \small uniCOIL $\rightarrow$ Contriever $\rightarrow$ ColBERTv2.
    \end{tablenotes}
    \end{threeparttable}
 	}
\end{table}
In Section~\ref{subsec:supervsion_strategies}, we mention that our sampling scheme treats top 10 passages from each teacher ranking list as positives and top 45--50 as negatives. 
We further conduct experiment to study the impact of the positive sampling scheme. 
Following the experiment setups in Section~\ref{subsec:empirical_study}, we use the sentences cropped from MS MARCO corpus as augmented queries and we conduct progressive label augmentation using top-$k$ passages as positive. 
The results are tabulated in Table~\ref{tb:positive_ablation}. 
We observe that treating top-$10$ passages from each teacher as positives yields the best supervised and zero-shot effectiveness. 
On the other hand, using only the top passage as positive results in significant effectiveness drop. 
This result indicates that the top passage labeled by a teacher cannot transfer its knowledge well to a student. 
This result is similar to the observation from \citet{spar}.

\subsection{An Intuition Behind Uniform and Progressive Supervisions}
\label{appedix:intuition}
As shown in Section~\ref{subsec:supervsion_strategies}, uniform supervision provides good supervision without fusion weight tuning as fused supervision.
Intuitively, a positive retrieved by more teachers has a higher probability to be sampled and may be more relevant to a query.   
To provide a sense of why uniform supervision works, we estimate the accuracy of supervision by computing the probability of each positives sampled under uniform supervision, and rank the positives according to the simulated probability. 
For example, at the 3rd iteration of progressive training, given a query, a positive passage is labeled positive by all the three teachers, the probability of the passages being sampled is $\frac{1}{3} \cdot ( \frac{1}{k} + \frac{1}{k} + \frac{1}{k}) = \frac{1}{k}$. 
In our experiments, each teacher labels the top 10 ($k=10$) retrieved passages as positives in our labeling scheme. 
Note that, in the case where multiple positives have equal probability, we further rank them according to their sum of reciprocal rank. 
For instance, if the two passages (e.g, $p_1$ and $p_2$) are retrieved by all the three teachers; then, we further rank them according to their scores $\frac{1}{r_{11}}+\frac{1}{r_{12}}+\frac{1}{r_{13}}$ and $\frac{1}{r_{21}}+\frac{1}{r_{22}}+\frac{1}{r_{23}}$, where $r_{mn}$ denotes the rank of the passage $p_m$ by the $n$-th teacher. 
In addition, we also estimate the diversity of supervision by computing the number of positive passages in union sets from the sources of supervisions.  

\begin{table}[t]
	\caption{Uniform and progressive supervision effectiveness comparison at each training iteration. The models are trained using cropped sentences as queries.}
	\label{tb:supervision_comparison}
	\centering
	 \resizebox{0.5\textwidth}{!}{  
    \begin{tabular}{l!{\color{lightgray}\vrule}ccc!{\color{lightgray}\vrule}ccc}
	\hline \hline
 \multirow{2}{*}{Teacher / iteration}& \multicolumn{3}{c}{uniform}& \multicolumn{3}{c}{progressive} \\
\cmidrule(lr){2-4} \cmidrule(lr){5-7} 
& 1& 2& 3& 1& 2& 3\\
\hline
uniCOIL & \checkmark& \checkmark & \checkmark& \checkmark& \checkmark& \checkmark\\
Contriever& \checkmark& \checkmark & \checkmark& \xmark& \checkmark& \checkmark\\
ColBERTv2& \checkmark& \checkmark& \checkmark& \xmark& \xmark& \checkmark\\
\arrayrulecolor{black}
\hline
MARCO Dev & 36.2& 37.0& 36.9& 34.9& 35.8& 36.6\\
 BEIR-13 & 46.6& 47.4& 47.6& 46.7& 48.6& 49.3\\
 \hline
 & \multicolumn{6}{c}{effectiveness of teacher} \\
  \arrayrulecolor{lightgray}
 \cline{1-7}
 MARCO Dev  & 39.1& 39.1& 39.1& 35.1& 36.5& 39.1\\
  \arrayrulecolor{black}
 \hline
 &\multicolumn{6}{c}{diversity of teacher} \\
  \arrayrulecolor{lightgray}
 \cline{1-7}
Avg. \# rel & 17.5& 17.5& 17.5& 10.0& 14.9& 17.5\\
\arrayrulecolor{black}
	\hline \hline
	\end{tabular}
	}
\end{table}

Table~\ref{tb:supervision_comparison} reports the detailed effectiveness and supervision quality (accuracy and diversity) comparison at each training iteration between uniform and progressive supervision as discussed in our pilot study. 
We observe that uniform supervision provides accurate and diverse supervision in the beginning of training; however, the generalization improvement over iteration is less than progressive supervision.

\begin{table}[h]
	\caption{MS MARCO and our augmented training queries statistics.}
	\label{tb:data_statistics}
	\vspace{-0.3cm}
	\centering
	 \resizebox{1\columnwidth}{!}{
	\setlength\tabcolsep{0.12cm}
    \begin{tabular}{l|r|r|r}
\hline \hline
  \multicolumn{1}{c}{}& \multicolumn{1}{c}{number}& \multicolumn{1}{c}{Avg. \# tokens}& \multicolumn{1}{c}{Avg. \# rel}\\ 
  \hline
  % \arrayrulecolor{lightgray}
passages in corpus& 8,841,823& 78.8& na\\ 
% \arrayrulecolor{black}
training queries& 532,761& 8.2& 1.0\\ \hline
\multicolumn{1}{c}{}&\multicolumn{3}{c}{\textbf{augmented training queries}} \\  \hline
cropped sentences& 28,545,938& 24.4& 23.1\\ 
generative queries& 28,545,938& 8.0& 24.7\\  \hline
\multicolumn{1}{c}{}&\multicolumn{3}{c}{\textbf{test queries}} \\  \hline
Dev& 6,980& 7.8& 1.1\\
DL19& 43& 7.6&95.4\\
DL20& 54& 7.5&68.0\\
\hline \hline
	\end{tabular}}
		\vspace{-0.4cm}
\end{table}

\subsection{MS MARCO Dataset Statistics}
\label{appedix:data_statistics}
Table~\ref{tb:data_statistics} lists the data statistics of MS MARCO dataset, including the original training queries and test queries (i.e., Dev, DL19 and DL20). 
In addition, we also list the augmented queries used to train \ourmodel s with full relevance label augmentation by five teachers.

\subsection{Latency Measurement for Relevance Label Augmentation}
\label{appedix:latency_measurement_for_rla}
We measure the latency of label augmentation using batch retrieval on a
single NVIDIA A100 40GB GPU for GPU search and 60 Intel(R) Xeon(R) Platinum 8275CL CPUs @ 3.00GHz for CPU search. 
For cross encoder, we conduct label augmentation by re-ranking text pairs with a batch size of 100. 
For dense retrieval (Contriever and GTR-XXL) and sparse retrieval, we use Faiss-GPU index and lucene index from Pyserini~\cite{pyserini} with 60 threads, respectively, and search with a batch size of 100. 
Note that we use a batch size of 25 to encode queries using GTR-XXL due to GPU memory constraint, which is also the main bottleneck for GTR-XXL batch retrieval.
For ColBERTv2, we use the improved version of multi-vector retrieval, PLAID~\cite{plaid}, and search with a batch size of 1, which is the only option.

\subsection{Model Checkpoints}
\label{appedix:model_checkpoints}

\smallskip \noindent \textbf{Teacher Models:} (1) uniCOIL: \url{https://huggingface.co/castorini/unicoil-msmarco-passage}; (2) Contriever: \url{https://huggingface.co/facebook/contriever-msmarco}; (3) GTR-XXL: \url{https://huggingface.co/sentence-transformers/gtr-t5-xxl}; (4) ColBERTv2: \url{https://github.com/stanford-futuredata/ColBERT}; (5) SPLADE++: \url{http://download-de.europe.naverlabs.com/Splade_Release_Jan22/splade_distil_CoCodenser_medium.tar.gz}; (6) Cross encoder: \url{https://huggingface.co/cross-encoder/ms-marco-MiniLM-L-12-v2}.

\smallskip \noindent \textbf{Baseline Models:} (1) CL-DRD: \url{https://github.com/HansiZeng/CL-DRD}; (2) RocketQAv2:\ we directly copy the numbers from \citet{colbert-v2}; (3) COT-MAE: \url{https://huggingface.co/caskcsg/cotmae_base_msmarco_retriever}; (4) Retro-MAE: \url{https://huggingface.co/Shitao/RetroMAE_BEIR}; (5) coCondenser: \url{https://huggingface.co/Luyu/co-condenser-marco-retriever}; (6) Contriever: \url{https://huggingface.co/facebook/contriever-msmarco}; (7) COCODR: \url{https://huggingface.co/OpenMatch/cocodr-base-msmarco}; (8) Promptgator (PTR) and GPL:\ we directly copy the numbers from their original papers~\cite{promptgator, gpl}.

\begin{table}[t]
	\caption{\ourmodel s' detailed effectiveness on LoTTE.}
	\label{tb:lotte_comparison}
	\centering
	 \resizebox{0.5\textwidth}{!}{  
	  	 \setlength\tabcolsep{5pt}
    \begin{tabular}{clcc!{\color{lightgray}\vrule}ccccccccccc}
	\hline \hline

& &sparse& multi-vec& \multicolumn{10}{c}{dense}\\ \hline
& & 0& 1& C& D& E& F\\
\hline
&&\rotatebox[origin=c]{290}{SPLADE++}&
 \rotatebox[origin=c]{290}{ColBERTv2}&
 \rotatebox[origin=c]{290}{\ourmodel-S}&
 \rotatebox[origin=c]{290}{\ourmodel-Q}&
 \rotatebox[origin=c]{290}{\ourmodel}& 
 \rotatebox[origin=c]{290}{\ourmodel+}\\
\hline
 \arrayrulecolor{black}
\hline
\multirow{5}{*}{\rotatebox[origin=c]{90}{\textbf{Search}}} &writing& 78.7& 80.1& 78.8& 78.2& 79.2& 81.5\\
& recreating& 71.9& 72.3& 73.4& 74.6& 76.0& 73.9\\
& science& 56.6& 56.7& 55.3& 56.9& 56.9& 57.9\\
& technology& 65.9& 66.1& 64.9& 68.8& 65.4& 67.6\\
& lifestyle& 83.7& 84.7& 84.9& 84.7& 85.6& 85.9\\
\arrayrulecolor{lightgray}
\hline
& pooled& 70.9& 71.6& 71.4& 72.4& 72.6& 73.5\\
\arrayrulecolor{black}
\hline
\arrayrulecolor{black}
\multirow{5}{*}{\rotatebox[origin=c]{90}{\textbf{Forum}}} &writing& 75.2& 76.3& 76.2& 75.2& 75.6& 77.5\\
& recreating& 69.2& 70.8& 69.9& 69.3& 70.3& 69.1\\
& science& 44.9& 46.1& 40.1& 40.1& 40.7& 41.4\\
& technology& 53.1& 53.6& 50.5& 51.2& 50.5& 51.4\\
& lifestyle& 76.9& 76.9& 77.0& 77.4& 76.9& 77.7\\
\arrayrulecolor{lightgray}
\hline
& pooled& 62.3& 63.4& 61.1& 61.2& 61.4& 62.1\\
\arrayrulecolor{black}
	\hline \hline
	\end{tabular}
	}
\end{table} 
\subsection{Detailed Effectiveness on LoTTE}
\label{appedix:lotte}
Table~\ref{tb:lotte_comparison} lists \ourmodel's more detailed effectiveness on five topics without aggregation. 
Although all the variants of \ourmodel show strong effectiveness on the datasets, we find that \ourmodel s perform poorly on the Forum queries about topics of science and technology compared to SPLADE++ and ColBERTv2. 
Combining science corpus pre-training and \ourmodel\ training may address the issue.

% listed in Table~\ref{tb:supervision_quality}. 

% In addition, progressive supervision is a variant of uniform supervision in order to provide dense retriever with easy-to-hard supervision.  % \subsection{Why Progressive Sampling Works?}

% \begin{figure}[t]
% \begin{subfigure}{0.49\columnwidth}
% \includegraphics[width=\columnwidth]{ms_trajectory.pdf}
% \caption{MS MARCO Dev}
% \end{subfigure}
% \begin{subfigure}{0.49\columnwidth}
% \includegraphics[width=\columnwidth]{beir_trajectory.pdf}
% \caption{BEIR-13}
% \end{subfigure}
% \caption{Effectiveness vs progressive training iteration.}
% \label{fig:sampling_trajectory}
% \end{figure}

\end{document}